\documentclass[preprint2, 10pt]{aastex}
\usepackage{amsmath}
\usepackage{graphicx}
\usepackage{bm}
\usepackage{epsfig}
\usepackage{amssymb,latexsym,mathrsfs}
\usepackage{graphicx}
\usepackage{color}
\usepackage{hyperref}
\usepackage{url} 

\usepackage[OT2,T1]{fontenc}
\DeclareSymbolFont{cyrletters}{OT2}{wncyr}{m}{n}
\DeclareMathSymbol{\Sha}{\mathalpha}{cyrletters}{"58}

\newcommand{\adsurl}[1]{\href{#1}{ADS}} 
\providecommand{\url}[1]{\href{#1}{#1}}

\newcommand{\be}{\begin{equation}}
\newcommand{\ee}{\end{equation}}
\newcommand{\bs}{\begin{split}} 
\newcommand{\bea}{\begin{eqnarray}}
\newcommand{\eea}{\end{eqnarray}}

\newcommand{\om}{\Omega_m}

\begin{document}
\title{Giving Cosmic Redshift Drift a Whirl} 
\author{Alex G.\ Kim\altaffilmark{1}, Eric V.\ Linder\altaffilmark{1,2}, Jerry Edelstein\altaffilmark{2}, David Erskine\altaffilmark{3}}
\altaffiltext{1}
{
Physics Division, Lawrence Berkeley National Laboratory, 1 Cyclotron Road, Berkeley, CA, 94720, USA}
\altaffiltext{2}
{
Space Sciences Laboratory, University of California Berkeley, 7 Gauss Way, Berkeley, CA 94720, USA
}
\altaffiltext{3}
{
Lawrence Livermore National Laboratory, Livermore,
CA 94550, USA
} 

\begin{abstract} 
Redshift drift provides a direct kinematic measurement of cosmic acceleration 
but it occurs with a characteristic time scale of a Hubble time.  Thus 
redshift observations with a challenging precision of $10^{-9}$ require a 
10 year time span to obtain a signal-to-noise of 1. We discuss theoretical 
and experimental approaches to address this challenge, potentially requiring 
less observer time and having greater immunity to common systematics.  On 
the theoretical side we explore allowing the universe, rather than the 
observer, to provide long time spans; speculative methods include radial 
baryon acoustic oscillations, cosmic pulsars, and strongly lensed quasars. 
On the experimental side, we explore beating down the redshift precision 
using differential interferometric techniques, including externally 
dispersed interferometers and spatial heterodyne spectroscopy. Low-redshift
emission line galaxies are identified as having high cosmology 
leverage and systematics control, with an 8 hour exposure on a 10-meter 
telescope (1000 hours of exposure on a 40-meter telescope) potentially 
capable of measuring the redshift of a galaxy to a precision 
of 
$10^{-8}$ 
(few $\times 10^{-10}$). Low-redshift redshift drift also has very strong 
complementarity with cosmic microwave background measurements, with the 
combination achieving a dark energy figure of merit of nearly 300 (1400) 
for 5\% (1\%) precision on drift. 
\end{abstract}


%\date{\today} 

%%%%%%%%%%%%%%%%%%%%%%%%%%%%%%%%%%%%%%%%%%%%%%%%%%%
\section{Introduction} \label{Sec:intro} 

Our universe is dynamic, i.e.\ the metric evolves, or more simply the 
scale factor of the universe changes with time: hence $a(t)$.  This 
gives rise to the cosmic redshift of light from distant sources.  Moreover, 
since the expansion rate itself evolves, in all but a coasting, Milne 
universe, the redshift of an object with fixed comoving coordinate will shift.  This redshift drift was 
introduced by \citet{sandage} and McVittie \citet{mcvittie} in the 
early 1960s and revisited in the 1980s and 1990s; see the textbook 
summary in \citet{fpoc}. 

Just as redshift is direct, kinematical evidence for cosmic expansion, 
redshift drift is likewise for cosmic acceleration.  This directness is 
an attractive feature as it does not depends on dynamics, i.e.\ the equations 
of motion (and separation of matter and dark energy density). 
Therefore, 
even if such a cosmic probe happens not to reach practically the same 
accuracy on dynamical cosmological-model parameters as more established
distance or growth of structure probes,  it is still worthwhile exploring possibilities 
for carrying it out. 

Conventionally, this is thought of in a brute force approach: stare at an 
object for a long time and measure the change in its redshift.  
Since the time scale for an order unity variation is the characteristic 
time scale of the expansion, the Hubble time of $\sim10^{10}$ y, this 
requires fantastically accurate measurements of redshift stable over the 
observing period.  Moreover, any other time variation in the metric that 
is not linear gives a competing effect.  From the Principle of Equivalence, 
any acceleration of the source or change in gravity along the line of sight 
acts as a systematic contribution. 

Here we explore two parallel tracks to make this cosmological probe more 
viable. One is exploring theoretical ideas for letting the universe do the 
difficult work for us, by taking advantage of source redshifts delivered 
to us at effectively different epochs and of differential rather than absolute 
measurements.  The second involves experimental approaches to improve the 
brute force precision, while again using differential measurements to 
ameliorate instrumental systematics. As with any probe, systematics are a 
key concern so any method that may enable better control of them is 
important to consider. 

In Sec.~\ref{sec:basic} we review the basic redshift drift and the plethora 
of systematics that confront it, as well as the potential leverage on 
cosmological model parameters.  Section~\ref{sec:methods} outlines three 
theoretical alternatives to the brute force approach, each with their own 
advantages and speculative aspects. 
We turn to some innovative experimental approaches in Sec.~\ref{sec:tech}, 
which may offer improvements in precision and systematics, and conclude in 
Sec.~\ref{sec:concl}.

%%%%%%%%%%%%%%%%%%%%%%%%%%%%%%%%%%%%%%%%% 
\section{Redshift Drift and Systematics} \label{sec:basic} 

In back-to-back articles published in 1962, \citet{mcvittie} and 
\citet{sandage} laid out the basics of redshift drift.  Since the 
cosmic redshift $z=a^{-1}-1$, then the dependence $a(t)$ of the source with 
respect to the observer necessarily implies that $z$ changes with time as 
well. Following the approach in \citet{fpoc}, we have 
\bea 
\frac{dz}{dt_0}&=&\frac{d}{dt_0}\left[\frac{a(t_0)}{a(t_e)}-1\right] 
=\frac{\dot a(t_0)-\dot a (t_e)}{a(t_e)}\\ 
&=&(1+z)\,H_0-H(z)\ , \label{zdot:eqn}
\eea 
where $t_0$ is the time the signal is observed in the observer-frame, 
$t_e$ the signal emission time, and $H=\dot a/a$ the Hubble expansion rate. 
Note that a dot superscript denotes a derivative with respect to the 
time argument shown. 

The form involving a difference 
between $\dot a$'s makes clear that this arises from acceleration (positive 
or negative).
Only a universe where $\dot a=\mbox{constant}$, i.e.\ $a\propto t$, 
has no redshift drift.  This is a Milne universe, which is conformal to a 
Minkowski spacetime, and so the Principle of Equivalence assures us that 
a universe without gravity has no acceleration, and vice versa. 
We emphasize the kinematic aspect of the redshift drift:
its nonzero 
value at any redshift directly indicates,
with no further assumptions about $a(t)$,
that the value of $\dot{a}$ differs at two different times and hence that there was an acceleration
(positive or negative).
In contrast, while distance measurements are kinematic, they must be differenced or 
differentiated to reveal acceleration (ideally through a regression 
method, e.g.\ \citet{2012PhRvD..85l3530S,2012JCAP...06..036S}), or fit to 
a dynamical model \citep{1998AJ....116.1009R,1999ApJ...517..565P}. 
Only when we want to compare 
to a particular model, e.g.\ so much matter density and so much dark energy 
density with some equation of state, do we need to know $a(t)$, i.e.\ the dynamics.

At high redshift we expect the universe to be decelerating, and hence 
the drift should be negative. 
At lower redshift, when the expansion rate $\dot{a}$ is lower than today and thus universe has experience acceleration, the drift will be
positive.
Thus the hope is to map out the 
influence of dark energy, and its equation of state, by accurate 
measurements of the redshift drift (in addition to establishing directly 
the mere presence of positive acceleration).

Figure~\ref{fig:sens} shows the sensitivity of the drift $\dot z$ as a 
function of source redshift to the cosmological parameters of the matter 
density $\om$
(scaled by a factor 20 so it can be seen over the full redshift range)
and dark energy equation of state $w(a)=w_0+w_a(1-a)$, where 
$w_0$ is its present value and $w_a$ a measure of its time variation.
The results
assume
a flat universe with
the fiducial values $\om=0.3$, $w_0=-1$, $w_a=0$ and the 
vertical scale is in units of $H_0$.  In these units, 
$\partial\dot z/\partial H_0$ is 
given by the $\dot z$ curve. The key properties to notice are: 
1) The sensitivity curves are different 
shapes, indicating no strong covariance between the effects of the different 
parameters, as long as measurements cover a range of redshifts, 
2) The greatest sensitivity is to the matter density, and this 
continues to outweigh the equation of state parameters at higher redshifts, 
by ever increasing factors, and 3) The initial rise of the equation of 
state sensitivities is sharp, achieving 50\% of the maximum sensitivity 
at quite low redshifts. These characteristics suggest that low-redshift 
measurements, 
{especially if, being closer and thus appearing brighter, the sources are more
easily observed to high signal-to-noise,
could be valuable for cosmological leverage on dark energy 
through the redshift drift probe. We test and quantify this below, and will see that this in fact fits in 
well with the new experimental techniques introduced in Sec.~\ref{sec:tech}.

%%%%%%%%%%%%%%%%%%%%%%%%%%%%%%%  
\begin{figure}[tbp] 
   \centering
   \includegraphics[width=\columnwidth]{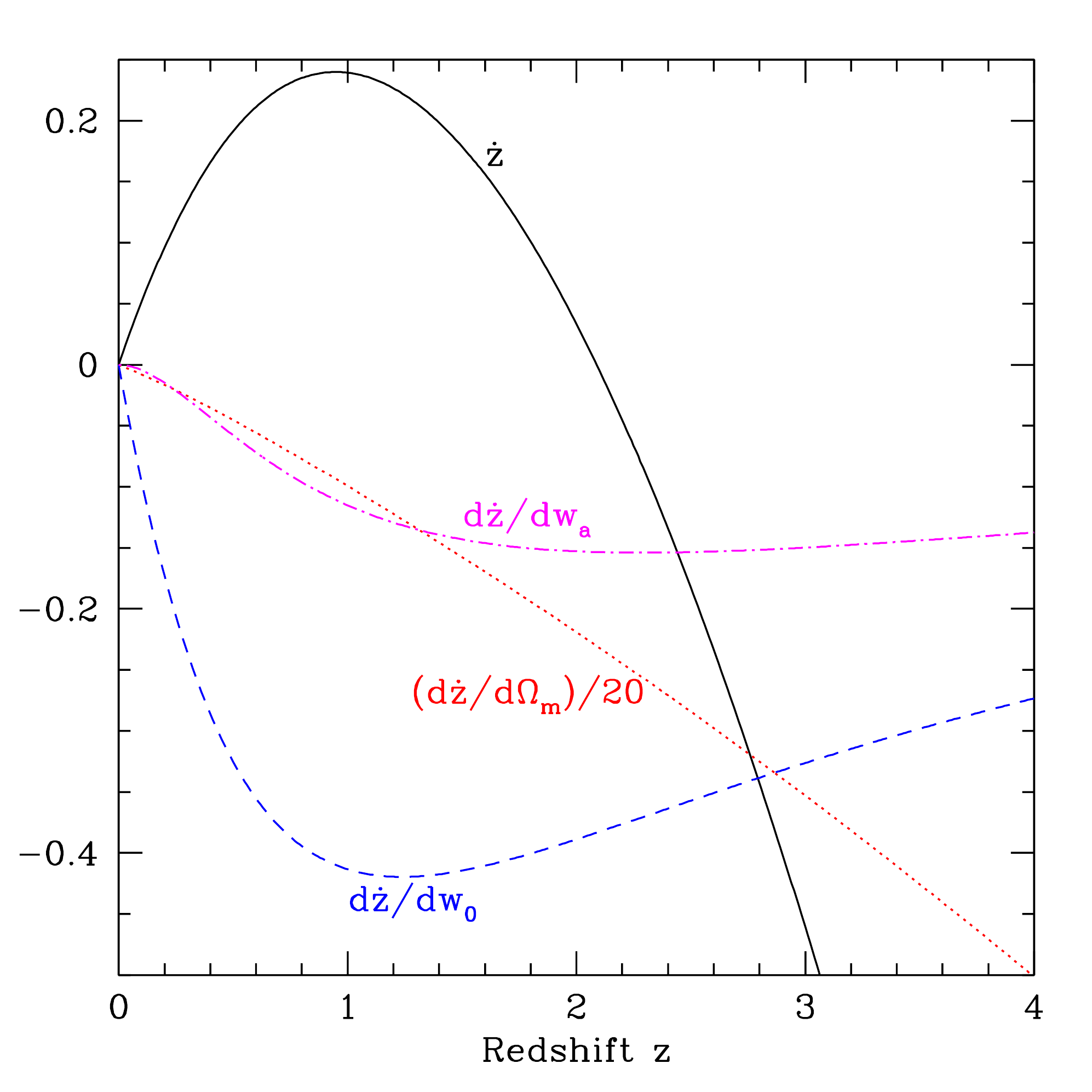}
\caption{The dependence of the redshift drift $\dot z$ and the sensitivities 
$\partial\dot z/\partial p$ to various cosmological parameters $p$ are shown 
as functions of redshift. The sensitivity with respect to $\Omega_m$ is 
scaled by a factor 20 to show its behavior over the redshift range. All quantities are in units of $H_0$, hence 
the redshift drift over an observing time $\Delta t$ is 
$\Delta z\lesssim H_0\Delta t=10^{-10}\,h(\Delta t/{\rm y})$. 
}
\label{fig:sens} 
\end{figure}

To first order, at
low redshift $\dot z\approx -H_0q_0z$ so the cosmological sensitivity 
initially increases linearly with redshift. 
In the photon-noise limit, the precision of a class of sources will go inversely with the luminosity distance. 
Since luminosity distance grows initially linearly with redshift, then 
more quickly at higher redshift, this indicates that low redshift may give 
the strongest cosmological leverage. 
At higher redshift, the sensitivity to matter density 
continues to increase, growing from linear with $z$ to 
proportional to $z^{3/2}$ at high redshift, swamping the dark energy 
parameters, whose sensitivities decline at high redshift. 
Precision at high redshift can only be recovered by considering different sources whose
brighter luminosities make up for the longer luminosity distance. 
All these 
characteristics give merit to further exploration of the use of 
low-redshift sources. 

From a theoretical 
sensitivity standpoint, without attempting a detailed observational 
strategy, we can quantify the redshift sensitivity. We carry out Fisher 
analysis using the sensitivities exhibited in Fig.~\ref{fig:sens} to 
explore the 
leverage on dark energy 
parameters $w_0$ and $w_a$, and their figure of merit (FOM) 
$[\det Cov(w_0,w_a)]^{-1/2}$ inversely proportional to the area of their joint 
uncertainty contour.

Figure~\ref{fig:fom} shows the results for the illustrative case of 
five measurements of $\dot z$ at 1\% precision each, centered on $z$, 
(the multiple measurements slightly spread 
in redshift are needed to allow fits of multiple cosmology parameters, 
while still focusing the sensitivity at a particular redshift z; specifically we use 
$z-0.2$, $z-0.1$, $z$, $z+0.1$, $z+0.2$, while 
see below 
for consideration of combining measurements at significantly different redshifts). We marginalize 
over $\om$, and over $H_0$ with a 
prior of $0.03$ (current precision) on 
$h=H_0/(100\,{\rm km\,s}^{-1}\,{\rm Mpc}^{-1})$. As expected from our previous discussion, 
the parameters are best determined from low-redshift observations, despite 
that $\dot z$ itself is largest at high redshift. This is due to a 
combination of dark energy being most prominent at low redshift, and 
that parameter covariances are smaller at low redshift. Note that near 
$z=2$, when $\dot z=0$ and so there is also no sensitivity to $H_0$, 
the leverage temporarily increases due to relief from covariance with 
$H_0$. However by $z\approx2.4$ the figure of merit has declined by more 
than a factor 10 from its low-redshift peak. This low-redshift leverage is 
advantageous since the signal-to-noise of the measurements we discuss later 
is much better at low redshifts. 

Figure~\ref{fig:fisher} illustrates the joint dark energy constraints 
from measurements at low, medium, and high redshift ranges, 
marginalizing over $\om$, and over $h$ with a prior of $0.03$. 
A further virtue 
of the low-redshift redshift drift probe is the positive orientation of 
the joint confidence contour, highly complementary to the negative orientation 
of nearly all other cosmological probes such as distances. This 
complementarity is lost for higher redshifts. However, if feasible, redshift 
drift measurements at high {\it and\/} low redshifts have great synergy. 
For example, the constraint from adding the $z=2.8$ measurements to the 
$z=0.3$ measurements yields a figure of merit of 784, about 6.6 times 
higher than the $z=0.3$ case alone. 
Optimization of survey redshift range, 
taking into account observation time constraints, is beyond the scope of 
this article, and is left for future work.

Alternatively, if we move away from 
purely kinematic observations and include measurement of the distance to 
cosmic microwave background (CMB) last scattering by Planck, then we obtain 
even greater synergy with low redshift drift observations -- a FOM of 1414, 
or 12 times higher than the $z=0.3$ case alone (also see the light grey 
FOM$_{\rm CMB}$ curve in Figure~\ref{fig:fom}). 
We have also investigated an alternate error model where 
instead of a constant fractional precision with redshift we have a constant 
absolute precision. This does not change the key aspect that low 
redshift still has the greatest leverage.

%%%%%%%%%%%%%%%%%%%%%%%%%%%%%%%  
\begin{figure}[!htbp] 
   \centering
  \includegraphics[width=\columnwidth]{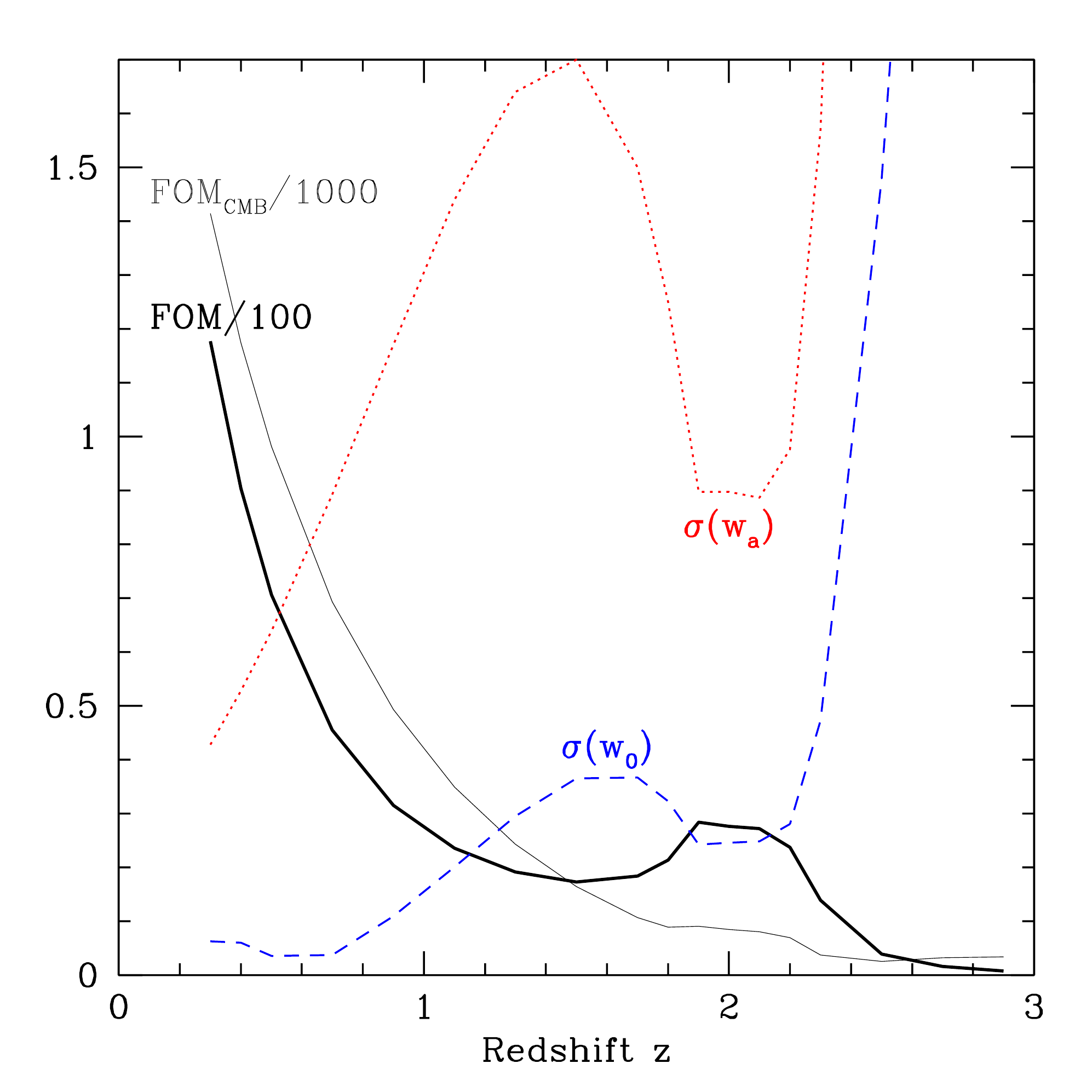}
\caption{Constraints at $1\sigma$ on $w_0$ and $w_a$, and their joint 
figure of merit (FOM), are plotted vs central redshift for experiments 
consisting of five measurements of redshift drift at 1\% precision. 
CMB constraints are included in (only) the FOM$_{\rm CMB}$ curve; 
note it is shown divided by 1000, rather than 100 like 
the FOM curve without CMB. 
} 
\label{fig:fom} 
\end{figure}

\begin{figure}[!htbp] 
   \centering
  \includegraphics[width=\columnwidth]{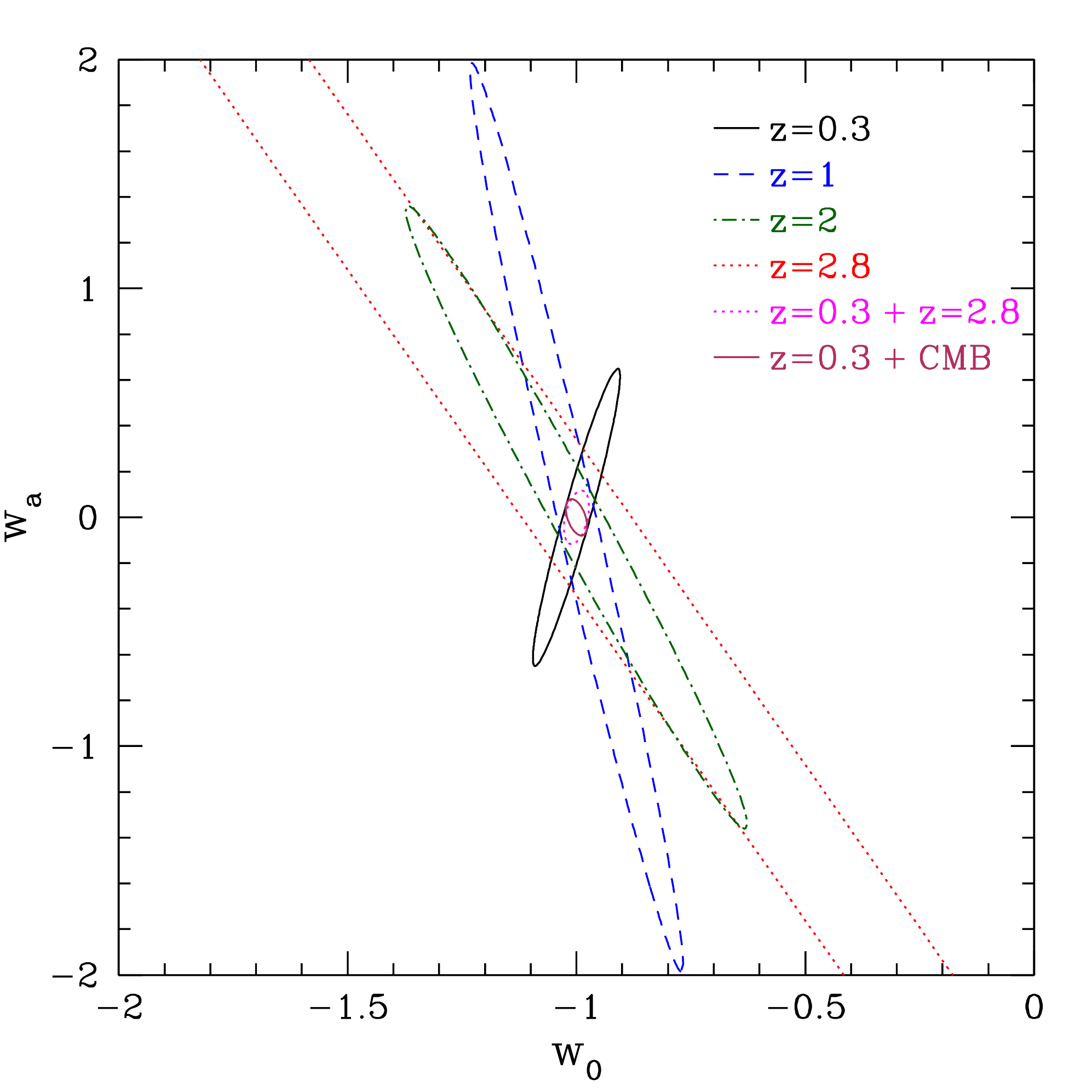}
\caption{Joint constraints on the dark energy equation of state parameters 
$w_0$ and $w_a$ are shown at 68\% confidence level, for experiments 
consisting of five measurements of redshift drift at 1\% centered at 
various redshift $z$. 
} 
\label{fig:fisher} 
\end{figure}

The key question, however, is how practical such measurements are, both 
with respect to needed precision and to systematic effects that degrade 
the accuracy. If the error floor lies at 5\% rather than 1\%, the $z=0.3$ 
case delivers only a figure of merit of 6.6, though the combined $z=0.3$ + 
CMB case gives 289, showing the strong complementarity. 
Figure~\ref{fig:fomprec} depicts how the figure of merit depends on the 
drift measurement precision; it stays above 100 (``stage 3'') out to 10\% 
precision. We also see that tightening the prior on $h$ helps considerably 
in the strong precision case, improving FOM by a factor 1.65 to over 2300 
when the prior goes from 0.03 to 0.01 (but for 5\% precision the factor is 
1.3).

%%%%%%%%%%%%%%%%%%%%%%%%%%%%%%%%%%% 
\begin{figure}[!htbp] 
   \centering
  \includegraphics[width=\columnwidth]{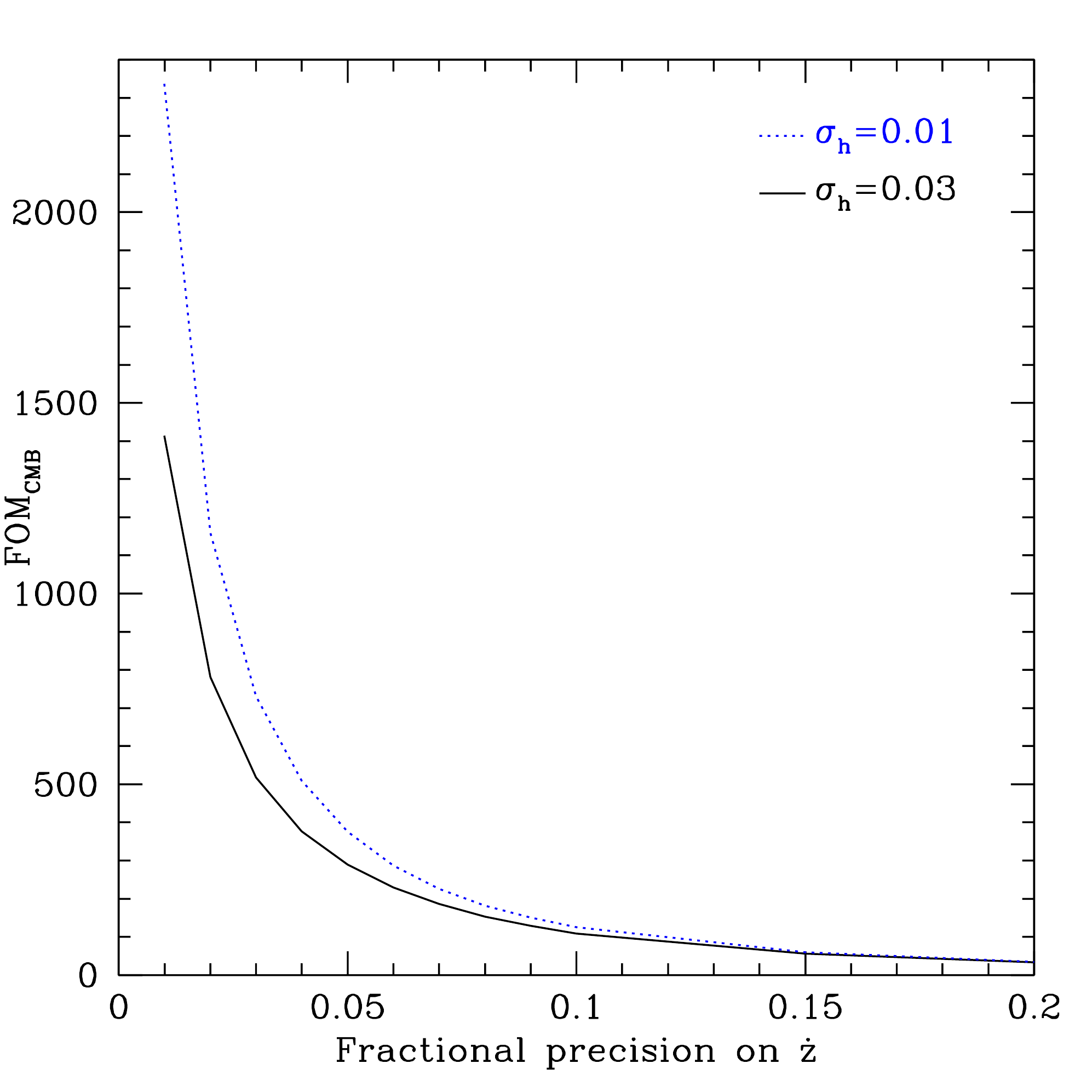}
\caption{The dark energy figure of merit is plotted as a function of 
redshift drift measurement precision, for a combination of five measurements 
centered at $z=0.3$ plus Planck CMB last scattering distance. The blue 
dotted curve shows the effect of tightening the Hubble constant prior to 
0.01. 
} 
\label{fig:fomprec} 
\end{figure}

We now outline 
some of the 
concerns regarding astrophysical systematic uncertainties. Sec.~\ref{sec:tech} 
will address improvements in precision and some experimental systematics 
using new experimental techniques. 

One astrophysical fly in the ointment is that in fact the observed redshift 
$z\ne a^{-1}-1$. The full relativistic expression is 
\be 
1+z=\frac{(g_{\mu\nu}k^\mu u^\nu)_e}{(g_{\mu\nu}k^\mu u^\nu)_o}\ , 
\label{eq:zgen} 
\ee 
with a subscript e denoting emitter and o denoting observer. 
The redshift drift obtains contributions from not only the 
homogeneous expansion in the metric, but 1) inhomogeneous gravitational 
potentials, 2) deviations in the photon four-momentum $k^\mu$ due to 
inhomogeneities, and 3) time variation of the source four-velocity $u^\nu$, 
i.e.\ peculiar accelerations.  

The full expression for the redshift drift including metric and geodesic 
effects but not peculiar acceleration is 
\bea 
\frac{dz}{dt_o}&=&\frac{\dot a_o-\dot a_e}{a_e}\\ 
&&+2[\dot\psi_e-(1+z)\dot\psi_o]
+\frac{2}{a_e}\partial_1(\psi_e-\psi_o)\nonumber\\ 
&&-(\dot\psi_e-\dot\phi_e)+(1+z)(\dot\psi_o-\dot\phi_o)\nonumber\\ 
&&-H(z)\,(\phi_o-\phi_e)+H_0\,(1+z)[a_o k^0_o]^{(1)}\ ,\nonumber  
\eea 
where $\psi$ and $\phi$ are the metric gravitational potentials, 
with a subscript e denoting emitter and o denoting observer. 
These terms are discussed in detail in \citet{10044646}; most are smaller 
than, though some are comparable to (especially for relativistic systems), 
peculiar accelerations. 

As for peculiar accelerations, using 
$\vec\nabla\psi\sim \dot u$, the contribution is 
\be 
\dot u \sim \frac{\psi}{L}\sim H\,\frac{H^{-1}}{L}\,\psi\sim H\, 
\left(\frac{40\,{\rm kpc}}{L}\right)\left(\frac{\psi}{10^{-5}}\right)\ , 
\ee 
where $L$ is a characteristic size of the system. 
For example, for a 
large galaxy of mass $10^{12}\,M_\odot$ with an emitting gas 
region at $L=10\,$kpc (at gravitational potential $\psi=5\times 10^{-6}$), 
the peculiar acceleration can be of order $2H$, an order of magnitude 
larger than the first, McVittie--Sandage term, and so must be reduced through 
the use of many independent sources, or sources in particularly ``quiet'' 
(smooth gravitational potential) regions.
See \citet{2008PhLB..660...81A,2008PhRvD..77b1301U} for 
more detailed calculations.
Note that when calculating the expected peculiar accelerations of astronomical sources one must recognize that those sources
preferentially occur in high-density regions, so that improper assumptions of the linearity of density perturbations or spatial averaging over the
power spectrum will generally underestimate peculiar accelerations.

Methods exist for ameliorating these systematics, e.g.\ observing several 
well separated sources 
with statistically independent peculiar accelerations. 
To be explicit, 
we recall that the precision required for measurement of 
redshift drift at a signal-to-noise of merely unity is 
$dz\approx 10^{-10}$ over a year (or $10^{-9}$ over 10 years), 
so a 1\% measurement would require $dz\approx 10^{-12}$ over a year.  
Systematics would need to be reduced below these levels.

%%%%%%%%%%%%%%%%%%%%%%%%%%%%%%%%%%%%%%%%% 
\section{Speculative Shortcuts for the Impatient} \label{sec:methods} 

Since the requirements on measuring $dz$ are so challenging, 
concepts for observational probes that inherently use long time scales or 
multiple measurements in one system would be of particular interest. 
A long lever arm in time and many measurements could reduce the requirements 
on individual measurement precision.  The universe itself provides sources 
at many different epochs, spread out over a Hubble time.  Likewise, 
individual sources can provide many different signals, e.g.\ spectral lines 
or wavetrains.  In this section we explore the use of each of these 
characteristics, and then their combination, in potential probes of 
acceleration. In general these either do not exactly give redshift drift, 
or are more speculative than established, so after this section we return 
to the stare-and-wait approach, but we introduce them to spur further thought.

%%%%%%%%%%%%%%%%%%%%%%%%%%%% 
\subsection{Many sources: Radial baryon acoustic oscillations} \label{sec:rbao} 
Measurement of the Hubble parameter, e.g.\ through the radial 
(redshift) extent of the baryon acoustic oscillation (BAO) scale, 
at different times is a kinematic
probe of the acceleration of the Universe. Just as redshift drift gives the difference
between $\dot{a}$ at two different cosmic times and its sign directly indicates
acceleration or deceleration, the difference in the Hubble parameter $\dot{a}/a$
at two different times gives an alternative measure of acceleration
and a different functional dependence on the cosmological parameters.
Here we briefly compare this Hubble drift from the 
familiar concept of radial baryon acoustic oscillations 
with redshift drift as discussed in the rest of this article.

We can think of a statistical ensemble of objects collectively as 
a source, for example 
defining a baryon acoustic feature in the density power spectrum.  This 
encodes a particular comoving scale, roughly the sound horizon $s$ at 
baryon-photon decoupling.  By measuring it along the line of sight, we 
are measuring a cosmic length $dr\sim150$ Mpc with endpoints at two 
different redshifts so $dr=dz/H$.  Since $dz\approx0.05$ then $H$ is 
fairly constant over this interval and we can simply say we obtain a 
measurement $H(z_1)$ from a slice at mean redshift $z_1$.  This is called a 
radial baryon acoustic oscillation (BAO) measurement, of the quantity 
$sH$. 

Carrying this out in a second redshift slice delivers $H(z_2)$.  Propagating 
the errors from each we can create a differential radial BAO (drBAO) 
measurement 
\be 
R=s\,(H_2-H_1) \ . \label{eq:drbao} 
\ee 
For redshift differences not small compared to unity, this measures the 
Hubble expansion parameter at times separated by a substantial fraction 
of the Hubble time, $\sim10^{10}$ y.  That is, we have left it to the universe 
to do the work of providing the long time baseline.  

While drBAO gives the ``Hubble drift'' rather than the redshift drift 
(in particular we measure the change in Hubble parameter over redshift, 
not the change in redshift over time), these two quantities are 
closely related, and indeed the Hubble drift measurement has some 
useful differences as shown in Table~\ref{tab:drbao}.  The nuisance 
parameter for comparing to cosmology theory is $H_0$ for the redshift 
drift, while it is the much better known sound horizon $s$ for drBAO.  
While redshift drift has a null where the expansion rate instantaneously 
matches a Milne coasting universe, 
drBAO has a null where the expansion rate matches a 
de Sitter universe. Finally, 
spectroscopic galaxy surveys measuring BAO also provide drBAO so no 
separate instrument or experiment is required.

%%%%%%%%%%%%%%%%%%%%%%%%%%%% 
\begin{table}[!htb]
%\footnotesize 
\small 
\begin{tabular}{l|ccc} 
Probe  &  Quantity  & Marg. & Sign Flip\\ 
\hline 
$z$ drift &$(1+z)\,H_0-H(z)$ & $H_0$ & Milne \\ 
drBAO & $s\,[H(z_2)-H(z_1)]$ & $s$ & de Sitter \\ 
\end{tabular} 
\caption{Redshift drift and differential radial BAO both 
directly measure the acceleration of the universe, but with different 
marginalization parameters and observational requirements. 
} 
\label{tab:drbao} 
\end{table} 
%%%%%%%%%%%%%%%%%%%%%%%%%%%% 
\subsection{Many signals: Cosmic pulsars} 

The long time baseline of the drBAO probe ameliorates the redshift 
precision needed, to simply that needed to measure the radial BAO scale 
($dz\approx10^{-3}$).  However it does involve the clustering of large scale 
structure and a statistical measure rather than simply photon propagation 
from an individual source.  If a single source gave many measurements, 
i.e.\ emitted many signals each of which could be used to measure redshift 
drift, then the requirement on the redshift precision can also be 
ameliorated in this manner.  
Conventionally this is thought of in terms of many spectral lines; we 
return to this later but here explore the time domain. 

If the source repeatedly emitted signals, then by measuring $N$ of them 
we could hope to statistically reduce the uncertainty on the redshift drift.  
Moreover, if there 
is a specific pattern to the signals, e.g.\ a periodicity, then further 
gains can be made.  This is the idea behind the period probe, or cosmic 
pulsar test; here we follow the textbook treatment of \citet{fpoc}.  The time 
observed for the arrival of the $N$th pulse with initial emitted period $P$ is 
\bea 
t_N&\equiv& t_o(NP)\nonumber\\ 
&=&NP\,\frac{dt_o}{dt_e}+\frac{1}{2}\,(NP)^2\,\frac{d^2t_o}{dt_e^2}+\dots\\ 
&=&NP\,(1+z)\nonumber\\ 
&\quad&+\frac{1}{2}\,(NP)^2\,(1+z)\left[(1+z)\,H_0-H(z)\right]\ . 
\eea 
The redshift drift (in square brackets) is thus enhanced by $N^2$ (that 
is, the precision on $\dot z$ scales as $N^{-2}$) and 
can be specifically fit to the quadratic behavior of the time series, 
helping to make a clean detection.  There is an extensive literature on 
pulsar timing and efficient fitting of different contributions 
\citep{pulsars}. 

While pulsars are fantastically regular clocks, well suited to this probe, 
we do not currently detect them at cosmological distances in the Hubble 
flow (but see \citet{13071628}; 
while this is for a transient, 
not periodic, source, an exciting prospect is that upcoming time domain surveys 
such as from LSST or SKA may find new classes of sources that could be used).
The idea behind cosmic pulsars is not 
restricted to neutron stars.  
One could consider other sources of (quasi-)periodic signals such as 
gravitational waves from supermassive binary black hole inspirals, which 
should be detectable 
at cosmic distances.  An issue here (other than detecting such sources) 
is that in these systems redshift is not measured separately but is tied 
with the black hole masses into the chirp mass combination.  Any evolution 
in the black hole masses -- or Newton's constant -- would be a systematic; 
see the analysis in Appendix B of \citet{09122724} following a suggestion 
by Linder.  Finally, any other accelerations, such as the motion of the system 
through an inhomogeneous gravitational potential or pulsar or gravitational 
wave kicks, would be a systematic.

%%%%%%%%%%%%%%%%%%%%%%%%%%%%%%%%%%%%%%%%%% 
\subsection{Multiple sources and signals: Strongly lensed quasars} \label{sec:lens} 

Strong gravitational lens images offer an observer distinct
benefits as a probe of redshift drift $dz/dt$. Strong lensing produces multiple images of the source, 
and so in a single system one can measure the redshift along multiple 
lines of sight, each having its own distinct time of photon flight (itself
a cosmological distance probe, see, e.g., \citet{13061272} 
and references therein).
The Universe gives us for free a time baseline for measuring
redshift drift of the individual source in a single moment of observer time.

Having systems with sufficiently long time delays to allow a
high signal-to-noise measurement of $\dot{z}$ in a single night would
be fantastic (although as will be discussed shortly, we cannot take advantage of intermediate redshift features from 
Lyman-$\alpha$ absorbers in long time delay lens systems with large angular image separation).
To date, the time delays measured range from days to
years \citep{2007ApJ...662...62F,2008ApJ...676..761F, 2013ApJ...764..186F}
and so have images with redshift
differences that are too small to detect in a single measurement with current methods.
Until longer-delay systems are identified, systems would
have to be monitored over a sufficiently long observer-time baseline.

Currently identified strong-lens systems have image angular separations from arcseconds
to under a half arcminute.
If the lines of sight are within the spectrograph field of view,
multiple images can be observed within a single exposure: sources of
wavelength calibration uncertainty that are correlated in that exposure
for the lines of sight cancel to reduce the contribution of wavelength
calibration as a source of uncertainty in the measurement of
wavelength differences.
For widely separated images, optical fibers can collect radiation from multiple images
and present them to the spectrograph for simultaneous exposures to obtain high velocity precision \citep{2000SPIE.4008..582P,2012SPIE.8446E..1VC}.

Multiple observations under different instrumental configurations
and observing conditions provide an important path toward reducing systematic
uncertainties. 

A strongly lensed variable source is used to get the time delay. A bright, time varying source 
such as an active galactic nucleus or quasar is ideal for this, and the time delay is measured from 
the photometric light curves (see, e.g.\ \citet{tewes,hkl} for statistical 
techniques for robust estimation).
Unlike for its use to obtain time-delay distances, 
we have no need to model the time delay through knowledge of the lens mass 
profile, image geometry, etc.
Since we do not need to model the lens mass profile, we can 
use cluster lenses, with larger time delays, rather than restricting to 
galaxy lenses as for the time-delay distance technique.

The spectra from two images of the same variable source observed on the same date
may not be the same.  
By the nature of the strong lensing time delay, it may be possible 
to cross-calibrate the spectra in both space and time.  That is, if the time 
delay between images A and B is one year, say, then when observing the system 
one year later the spectrum of image B should match that of image A from 
the previous year (modulo the redshift drift itself). 

We now consider the case of a strongly lensed quasar.  The time delay can
be measured as lags in the image light curves.  The quasar Q1101-264  has
embedded within its observed spectrum the redshifts of individual Lyman-$\alpha$ absorbers
and $225$ narrow spectral lines of metal absorbers along the light of sight;
the redshift drift for each can 
be measured 
with observations of a single image
over an extended time period
\citep{1998ApJ...499L.111L,2008MNRAS.386.1192L}.

It may be possible to use 
foreground Lyman-$\alpha$ absorption along the lines of sight to the multiple images 
of a strongly lensed quasar as additional indicators of redshift drift, 
but this entails certain complications.
The geodesics may not pass through the same gas clouds.
The transverse distance at redshift $z$ between 
lines of sight separated by angle $\Delta\theta$ is 
\be 
\Delta r=D_A\,\Delta\theta\approx 
6\left(\frac{H_0D_A}{0.4}\right)\left(\frac{\Delta\theta}{1''}\right)\, 
h^{-1}{\rm kpc} \ . 
\ee 
Note that for $z=1-3.5$, $H_0D_A$ is almost constant at 0.4 in a 
cosmology near the concordance model.  Image separations tend to be in 
the 1--$5''$ range.  For such transverse separations, much smaller than 
the gas Jeans scale of $\sim300\,h^{-1}$ kpc, the Lyman-$\alpha$ lines 
should be coherent in the multiple images \citep{09100250}.  
Very long time-delay systems have larger image separations, roughly 
$\Delta t\sim(\Delta\theta)^2$ and the coherence will be degraded for 
systems with separations greater than tens of arcseconds. 
The time delay for each gas cloud will not be directly measured, however they
can be (at least approximately) constrained using the delay
of the background quasar and the Friedmann-Robertson-Walker geometry. 

Wide-field surveys such as Dark Energy Survey and that planned for LSST will find 
approximately $10^3$ and $10^4$ strongly lensed quasars, respectively.  
Monitoring 
campaigns such as COSMOGRAIL \citep{cosmograil} and STRIDES \citep{strides} 
obtain long term light curves for the images, measuring the time 
delays.  Depending on the time delay, photometry quality, and other 
factors it may take $\sim$1000 nights to obtain a time delay 
estimation, though note that the accuracy requirements may be relaxed 
from the time delay distance case, e.g.\ 5\% accuracy may be sufficient 
for this not to contribute significantly to the redshift drift uncertainty. 
The differential redshift measurement itself takes only a single night, 
so the story of the universe may be revealed in 1000 nights and a night 
\citep{1001}.

Lensing can affect the image redshifts if the lens is moving 
\citep{mitrofanov,birkgull}.  From 
Eq.~(\ref{eq:zgen}) we see this arises because the dot product between 
the photon four-momentum and the source and observer four-velocities 
has to go through the intermediate step of the lens plane, where the 
deflection shifts the angle between the vectors.  The redshift contribution 
is of order $v\,\delta\alpha$, where $\delta\alpha\sim\psi$ is the 
deflection angle.  The redshift drift contribution is then 
$d(v\,\delta\alpha)/dt$, much smaller than a standard peculiar 
acceleration of the source and so can be neglected.

%%%%%%%%%%%%%%%%%%%%%%%%%%%%%%%%%%%%%%%%% 
\section{Experimental Approaches for High-Precision Redshift Drift Measurements} \label{sec:tech} 

In this section we turn from theoretical speculation to consideration of 
innovative experimental approaches, with practical details on how to 
obtain highly precise redshift measurements. 
To provide context, the redshift measurement with an uncertainty as low as 
$\sigma_z=2\times 10^{-8}$ for the radio absorption line system 0738+313A at $z=0.091$
\citep{2012ApJ...761L..26D} represents current high-precision 
work.
We are guided by two concepts: 
that low-redshift measurements can have significant cosmological leverage, 
and that the differential measurement of a single-epoch redshift from the wavelength separation
of line doublets can be more accurate than a measurement based on a single line.

An important point is not just precision but instrumental accuracy, 
e.g.\ stable calibration. 
Traditionally, for redshift drift one might wavelength calibrate a spectrum with an external 
standard, such as an iodine cell, and then repeat this ten or so years 
later with the next redshift measurement of the source.  However, the 
wavelength calibration is not guaranteed to remain fixed over this long time 
span; indeed \citet{griest1,griest2} find it can shift significantly over a 
single night with changes in temperature, humidity, etc.  
These shortcomings have led to 
laser frequency combs being the standard to achieve accurate wavelength calibration
\citep{2007MNRAS.380..839M,2008Sci...321.1335S,2012Natur.485..611W}.

%%%%%%%%%%%%%%%%%%%%%%%% 
\subsection{Emission Line Galaxies as Targets} 

Emission line galaxies are excellent candidates for the measurement of 
precision redshifts. The best sources involve bright, narrow lines that 
are clearly identifiable. 
The unique signature
of doublets amidst  other emission lines allows unambiguous identification of the [OII] $\lambda\lambda$3727--3729\AA\ 
and [OIII] $\lambda\lambda$4959--5007\AA\ doublets.
Their high line fluxes provide strong signals that suppress statistical Poisson uncertainty.
Doublets occupy only a narrow bandwidth that spectrographs must span; 
regions outside the doublet need not be measured. 
The two lines composing a doublet are produced by the same atoms and therefore
share a common line profile.
Also, emission line galaxies occur more in the field than in clusters, potentially ameliorating peculiar accelerations.

The wavelength separation of the doublet lines is proportional to $(1+z)$, 
so their cross-correlation in log-wavelength
thus provides a measure of redshift that is independent of the specific shape of the line profile.
The true functional forms of the line profiles are a priori unknown, which 
could otherwise lead to redshift uncertainty.  Quoted line effective-wavelength uncertainties
are generally up to a factor 10 finer than the resolution: for example a line with $\Delta v=3$~km\,s$^{-1}$ is measured to
precision $\Delta \lambda/\lambda \gtrsim 10^{-6}$.

The redshift drift measurement does not require knowledge of the effective wavelength but rather only the wavelength shift
of the profile, whatever the profile is. 
As just noted, within a single exposure the cross-correlation of the two doublet lines provides a robust redshift measurement.
Similarly, cross-correlation of multiple spectra taken at different epochs provides a robust redshift-drift measurement.

We study potential targets from the Sloan Digital Sky Survey data release SDSS3 DR10. 
The emission-line fits to the spectra described in \citet{2012AJ....144..144B} are used to select
objects 
with sharp and bright [OII] and/or [OIII] emission lines.
The redshifts, velocity dispersions $\Delta v$, line strengths, and host continuum levels for a select set of galaxies are shown in Table~\ref{lines:tab}. 
The best-fit forbidden line $\Delta v$ and the average continuum flux for each doublet  are given.
For target Plate~\#2872, Fiber~\#468, at $z=0.680$, the catalog does not provide significant measurements of [OIII] features.
The SDSS spectrum from a representative 
target, from Plate~\#2510, Fiber~\#560, is shown in Figure~\ref{shsinput:fig}.

%%%%%%%%%%%%%%%%%%%%%%%%%%%%%%%%%%%%% 
\begin{deluxetable}{ccccccccc}
\tablecolumns{9} 
\tabletypesize{\small}
\tablewidth{0pc} 
\tablecaption{Target Line Properties.\label{lines:tab}}
\tablehead{
\colhead{} &\colhead{} &\colhead{}&\colhead{} &\colhead{$\Delta v$}&\colhead{}   & \colhead{Flux 1}  & \colhead{Flux 2} & \colhead{Continuum}\\
\colhead{Plate} & \colhead{MJD} & \colhead{Fiber} &\colhead{$z$}&\colhead{(km\,s$^{-1}$)}&\colhead{Doublet} & \colhead{(erg\,s$^{-1}$cm$^{-2}$)}& \colhead{(erg\,s$^{-1}$cm$^{-2}$)}& \colhead{(erg\,s$^{-1}$cm$^{-2}$\AA$^{-1}$)}
}
\startdata
\tableline
2510 & 53877 & 560 &  0.045 & 75.510 &[OII] &$2.01\times10^{-14}$ &$2.15\times10^{-14}$ &$4.39\times10^{-16}$\\
&&&&
&[OIII] &$2.09\times10^{-13}$ &$6.92\times10^{-14}$ &$2.23\times10^{-16}$ \\ 
\tableline
1314 & 52792 & 303 &  0.150 & 36.940 &[OII] &$3.51\times10^{-17}$ &$6.17\times10^{-17}$ &$5.09\times10^{-17}$\\
&&&&
&[OIII] &$7.73\times10^{-15}$ &$2.55\times10^{-15}$ &$8.37\times10^{-17}$ \\ 
\tableline
594 & 52027 & 516 &  0.210 & 93.640 &[OII] &$4.41\times10^{-15}$ &$4.52\times10^{-15}$ &$5.97\times10^{-17}$\\
&&&&
&[OIII] &$3.28\times10^{-14}$ &$1.08\times10^{-14}$ &$3.45\times10^{-17}$ \\ 
\tableline
2872 & 54468 & 468 &  0.680 & 42.470 &[OII] &$3.44\times10^{-17}$ &$3.58\times10^{-15}$ &$1.18\times10^{-16}$\\
\tableline
\enddata
\end{deluxetable}

%%%%%%%%%%%%%%%%%%%%%%%%%%%%%% 
\begin{figure}[t]
   \centering
    \includegraphics[width=\columnwidth]{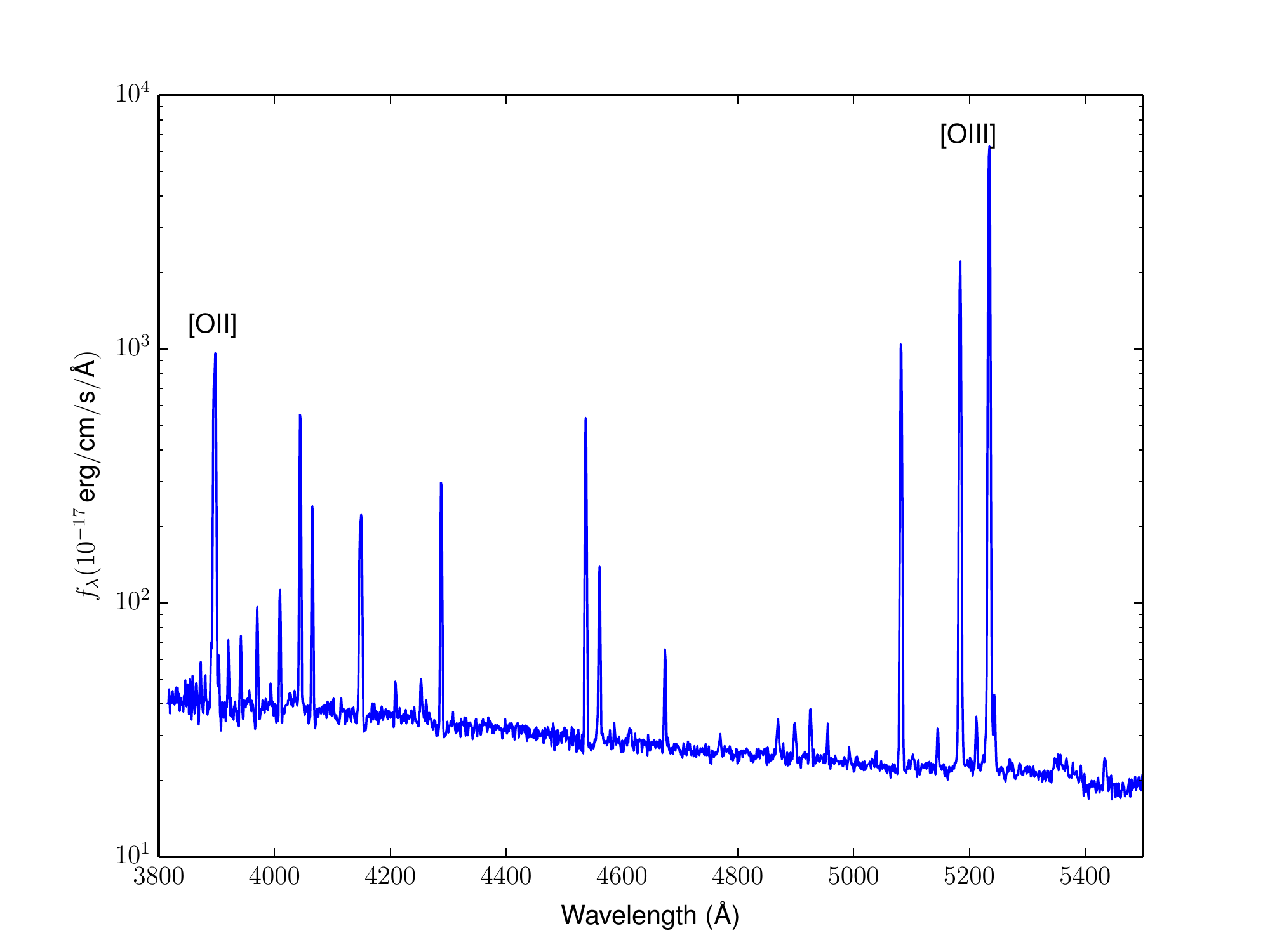} 
   \caption{Spectrum of SDSS target specified by Plate~\#2510, Fiber~\#560 from the SDSS3 DR10.    \label{shsinput:fig}}
\end{figure}

The tabulated galaxies are not the product of an exhaustive search and are meant to represent a minimum of what can be available
as good
candidate targets.
Future surveys such as eBOSS, Dark Energy Spectroscopic Instrument, and Euclid will specifically target emission-line
galaxies from which the best targets can be identified.
In the scenarios considered in this article, redshift uncertainty scales
with the 
line width and the inverse square root of line flux; smaller numbers 
for the precision are better so bright, narrow lines 
are desired. 

The Sloan spectrograph has only moderate resolution, which sets a floor on the
sharpness of the lines it can measure and the precision of the $\Delta v$ estimates.
Therefore, targets initially identified with $R\lesssim 2000$ spectrographs would benefit from subsequent screening with
high spectral-resolution, preferably integral field unit, spectrographs, in order to accurately determine which candidates
are best to pursue.

The spectral regions of interest
are around each doublet.  In the calculations that follow,
the area around two emission lines
is modeled as a function of wavenumber $\sigma$ as
\begin{align}
B(\sigma)& =\frac{a_1}{\sqrt{2\pi s_1^2}}\exp{\left[-\frac{\left(\sigma-\sigma_1\right)^2}{2s_1^2}\right]} \nonumber\\
 &\qquad +\frac{a_2}{\sqrt{2\pi s_2^2}}\exp{\left[-\frac{\left(\sigma-\sigma_2\right)^2}{2s_2^2}\right]}.
\label{input:eqn}
\end{align}
The two lines have central wavenumbers $\sigma_1$ and $\sigma_2$, and share a common Gaussian profile with width parametrized by velocity dispersion
such that $s_i=\sigma_i\Delta v/c$.
The value of 
$\Delta v$ is the same for the lines of a doublet.

%%%%%%%%%%%%%%%%%%%%%%%%%%%%%%%%%%%%%%%%%%%%%%%%%%%%%% 
\subsection{Spectrograph Types for Measurements} 

The precision with which redshift can be determined depends on the instrument used for the measurement.
We consider several spectrometers designs, including a ``Conventional'' high-resolution dispersion spectrograph and the
interferometric instruments ``EDI'',
``SHS'', and ``ED-SHS'' described in the following subsections.  Interferometers produce a Fourier transform of a signal, converting sharp
spectral features into wave patterns whose frequency is dependent on the feature wavelength.  Interferometers provide
increased statistical sensitivity to wavelength measurements and potentially have less sensitivity to systematic
uncertainties.

Some common assumptions are made for our analysis of spectrometer 
measurements. 
The baseline observation is for an 8-hour exposure on a 10-m telescope.
To allow direct comparison of the 
designs, all systems are assigned the same total throughput of
 35\%.
The interferometer optics and fringe visibility has a $\sim 75\%$ efficiency, so the Conventional spectrograph
will have a third better throughput relative to the other systems.

The dispersion spectrometer used for Conventional spectrograph has a two-pixel resolution $R=50{,}000$
whereas the EDI, and ED-SHS instruments is taken to have resolution $R=20{,}000$.  The resolution of the Conventional spectrograph is chosen in order to sample the features.
EDI and ED-SHS allow the use of lower resolution spectrographs (much cheaper and easier to make thermally stable)
because it is the interferometer component which measures the velocity.

For all cases considered the photon noise dominate uncertainties;
we include a detector read noise of $2e^-$, total integrations split into 2-hour exposures, and a dark current of $2e^-$\,h$^{-1}$,
which affect the calculated signal-to-noise at the least significant quoted digit.
Blocking filters remove flux (and hence noise) in wavelengths outside the regions
of interest.  The new moon CTIO sky emission is included and
is small compared to the galaxy continuum.

Projected redshift precisions are calculated with a Fisher matrix
analyses with the source redshift $z$ as the only free parameter.  The
equation for the predicted signal is given for each spectrograph type.
For conciseness, trivial behavior along the spatial axis is not given
explicitly in these equations.  Measurement uncertainties come from
photon and detector noise.
Uncertainties scale as the inverse square root of exposure time and inverse of telescope aperture. 
For the SHS, and when the other spectrographs resolve the spectral line, the precision scales 
inversely 
with the line dispersion velocity.  
As the measurements
are source-noise dominated, redshift precisions scale as the inverse square 
root
of line flux.

We are interested in uncertainties in redshift drift.
This is in some ways simpler than measuring absolute redshifts
because correlated redshift uncertainties cancel each other when measuring drift.
Instruments contribute sources of measurement uncertainty in redshift drift through 
uncertainty in the conversion from pixel counts to physical flux, uncertainty in the point spread function,
and limitations in the calibration
of wavelength.   These uncertainties distort intrinsic line shapes and so introduce error in the cross-correlation
of the doublet lines and spectra taken on different dates. 
Each spectrograph type
has different susceptibilities to systematic uncertainties.

%%%%%%%%%%%%%%%%%%%%%%%% 
\subsection{Conventional Dispersion Spectrograph} \label{sec:dispspec} 

The redshift of a galaxy can be measured from the output of a dispersion spectrograph, such as shown in Figure~\ref{shsinput:fig}.
For an input spectrum $B(\sigma)$ the expected signal is
\begin{equation}
I(\sigma) = \left[B(\sigma)\otimes \mbox{PSF}(\sigma)\right]\Sha\left(\frac{\sigma}{p}\right),
\label{conventional:eqn}
\end{equation}
where $p$ is the spacing of the wavenumber sampling and the Shah function $\Sha$ is the set of delta functions
that specify the discrete sampling.

The wavelengths of the observed lines are compared to the corresponding known restframe wavelengths to give
the redshift.
Projected statistical uncertainties from an $R=50,000$ spectrograph for the target galaxies are given under
the ``Conventional'' column of Table~\ref{dz:tab}.  Results from [OII], [OIII], and their combination are listed separately.

%%%%%%%%%%%%%%%%%%%%%%%%%%%%%%%
\begin{deluxetable}{ccccccc}
\tablecolumns{7} 
\tablewidth{0pc} 
\tablecaption{Statistical Redshift Uncertainties of Select SDSS Targets With
an 8 Hour Exposure on a 10-m Telescope For Different Spectrograph Types.
\label{dz:tab}}
\tablehead{
\colhead{Plate} &\colhead{Fiber}  &\colhead{Doublet}& \colhead{Conventional} & \colhead{EDI} & \colhead{SHS} &\colhead{ED-SHS}
}
\startdata
\tableline
2510 & 560 
& OII & $1.5\times10^{-7}$  & $4.9\times10^{-8}$  & $1.7\times10^{-7}$  & $4.8\times10^{-8}$  \\
& &OIII  & $3.5\times10^{-8}$  & $1.1\times10^{-8}$  & $4.3\times10^{-8}$  & $1.1\times10^{-8}$  \\
& &OII\&OIII  & $3.4\times10^{-8}$  & $1.1\times10^{-8}$  & $4.2\times10^{-8}$  & $1.1\times10^{-8}$  \\
\tableline
1314 & 303 
& OII & $2.5\times10^{-6}$  & $7.6\times10^{-7}$  & $2.8\times10^{-6}$  & $8.5\times10^{-7}$  \\
& &OIII  & $1.0\times10^{-7}$  & $3.2\times10^{-8}$  & $1.4\times10^{-7}$  & $3.2\times10^{-8}$  \\
& &OII\&OIII  & $1.0\times10^{-7}$  & $3.2\times10^{-8}$  & $1.4\times10^{-7}$  & $3.2\times10^{-8}$  \\
\tableline
594 & 516 
& OII & $4.6\times10^{-7}$  & $1.5\times10^{-7}$  & $7.2\times10^{-7}$  & $1.5\times10^{-7}$  \\
& &OIII  & $1.2\times10^{-7}$  & $3.8\times10^{-8}$  & $1.4\times10^{-7}$  & $3.8\times10^{-8}$  \\
& &OII\&OIII  & $1.2\times10^{-7}$  & $3.7\times10^{-8}$  & $1.4\times10^{-7}$  & $3.7\times10^{-8}$  \\
\tableline
2872 & 468 
& OII & $2.9\times10^{-7}$  & $9.2\times10^{-8}$  & $2.7\times10^{-7}$  & $9.3\times10^{-8}$  \\
\tableline

\enddata
\end{deluxetable}

Not included in Table~\ref{dz:tab} are sources of the instrumental systematics.
Wavelength calibration is performed through observations of arc lamps or laser frequency combs
emitting lines at known wavelengths.
 If the arc is taken simultaneously
with the science exposure, a minimum wavelength interpolation distance must be maintained so that the arc does
not interfere with the doublet.  Otherwise, the arc must be observed
spatially offset from the science signal on the detector or in a different exposure.
Therefore,  temporal, spatial, and/or
wavelength interpolation are applied to calibrate wavelengths.

Uncertainty in the often variable and charge-dependent point spread function can
bias the determination of the line centroids.  Just as with the flux calibration requirement, sub-per-mil accuracy in predicted pixel counts
is required.  For redshift drift, these calibration requirements are differential between observations taken
over the duration of the survey: if absolute calibration is not achieved, instrumental stability is essential.

%%%%%%%%%%%%%%%%%%%%%%%%%%%%%%%%%%%%% 
\subsection{Externally Dispersed Interferometer} 

The Externally Dispersed Interferometer \citep[EDI;][]{2003PASP..115..255E} is a candidate instrument to measure precision redshifts. 
An EDI is the sequence of a Fourier transform spectrograph (FTS) and a dispersion spectrograph.  The FTS is an interferometer that shifts
the phase of incoming coherent light
by an amount dependent on wavelength and a  delay between the two arm lengths.
The ensuing dispersion spectrograph takes the phased light and separates it into fine wavelength bins.
For an individual wavelength bin the output signal depends on the phase introduced by the interferometer; multiple
measurements taken after dithering the delay  make apparent a modulation in the output signals.
Therefore, in an EDI the wavelength can be measured using both the calibration methods of a standard dispersion
spectrograph {\it and\/} from the modulations of signal apparent when changing the FTS arm-lengths.   As a consequence,
it was shown by
\citet{2003PASP..115..255E} that EDI's provide line-velocity measurements that are more precise than with a dispersion spectrograph. 
Basically, the sinusoidal interferometer modulations 
provide a steeply sloped PSF that responds sensitively to changes in 
line position. This steep slope improves the effective resolution, as 
illustrated in Fig.~\ref{fig:edipsf} and gives 
higher sensitivity in redshift measurement.

%%%%%%%%%%%%%%%%%%%%%%% 
\begin{figure}[t]
   \centering
   \plotone{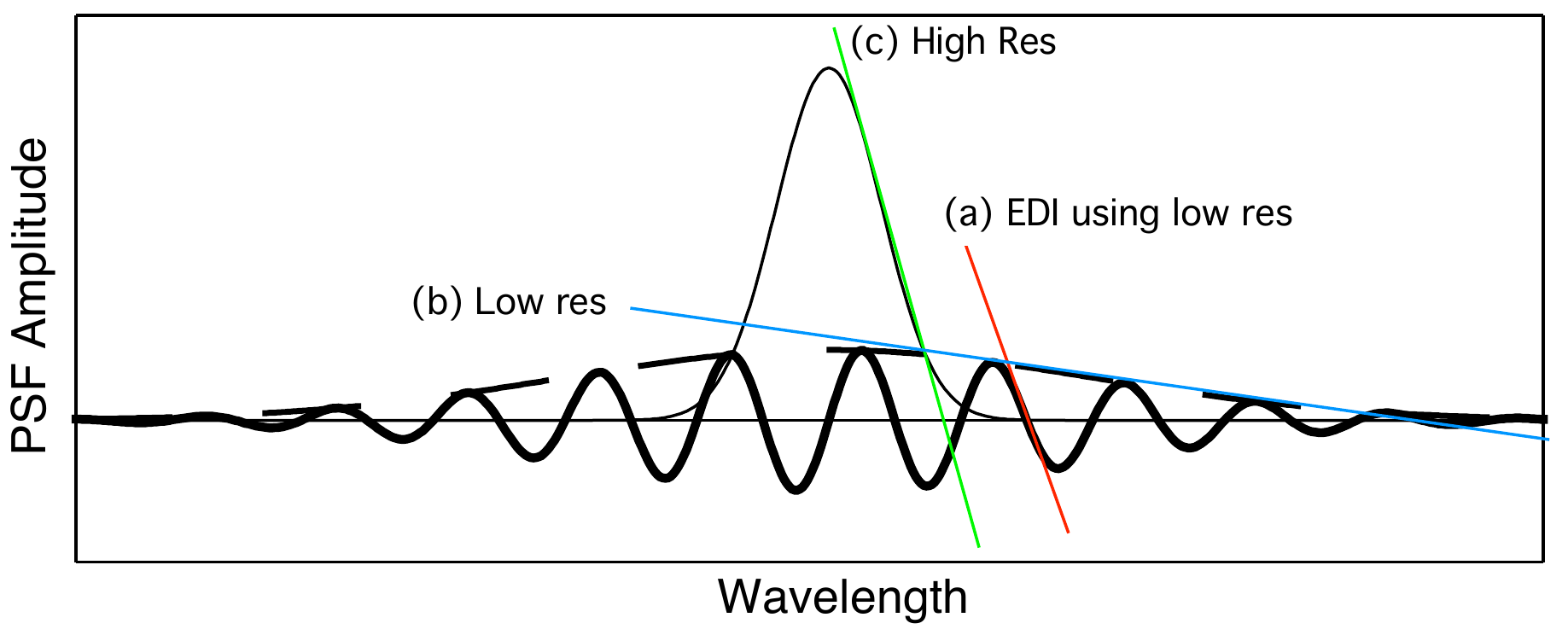} 
   \caption{An illustrated view showing that 
the PSF of an EDI (a) using a low resolution spectrograph (b) has steep slopes similar to a high resolution spectrograph (c), because the EDI uses an interferometer to distinguish fine wavelength differences.  Hence the PSF of an EDI has sinusoidal corrugations of user selectable period, which is optimally set to about twice the instrinsic feature linewidth.  The Doppler sensitivity is largely controlled by the PSF slope, so the EDI can achieve reasonably high Doppler sensitivities in spite of using a dispersive spectrograph that may be too low, by itself, to resolve a spectral line. 
\label{fig:edipsf}}
\end{figure}

An EDI produces measurements that can be related
to a conventional spectrum
\begin{equation}
I(\sigma) = B(\sigma) \otimes \mbox{PSF}(\sigma),
\end{equation}
and an independently measurable whirl
\begin{equation}
W(\sigma) = \frac{1}{2}\left[e^{i2\pi\tau\sigma}B(\sigma)\right] \otimes \mbox{PSF}(\sigma),
\end{equation}
where $B$ is the input spectrum, $\tau$ is the interferometer delay
with units of distance, and $\mbox{PSF}$  is the blurring
response of a pure frequency.

The heterodyning feature of the EDI is seen in the Fourier transforms of the above expressions:
\begin{align}
i(\rho) & =b(\rho) \, \mbox{psf}(\rho)\\
w(\rho -\tau) & =\frac{1}{2}\, b(\rho) \, \mbox{psf}(\rho-\tau).
\end{align}
The EDI whirl has the instrumental frequency response shifted into a new $\rho$ range to allow
spectral
resolution on scales tuned by the choice of $\tau$.  In the calculations that follow, the delay is chosen to resolve
the FWHM of the line with the lower wavenumber, $\tau = (2.36s_1)^{-1}$, where $s_1$ 
is the standard deviation width of the first line as 
in Eq.~(\ref{input:eqn}). 

An EDI observation is actually a series of measurements
\begin{align}
I(\sigma)_{\Delta\tau} &= \left[B(\sigma)\left(1+\cos{\left(2\pi\left(\tau+\Delta \tau\right)\sigma+\phi_y \right)} \right) \right. \nonumber\\
&\qquad \left. \otimes \mbox{PSF}(\sigma)\right]\Sha\left(\frac{\sigma}{p}\right),
\label{edisig:eqn}
\end{align}
where $p$ is the spacing of the wavenumber sampling, $\phi_y$ is the initial phase,
and $\Delta \tau$ represents the changes in delay in the series of exposures; for our calculations we choose  $\Delta \tau = (n/4)[2/(\sigma_1+\sigma_2)]$ with $n\in\{0,1,2,3\}$, 
achieved by adjusting an interferometer arm length by quarters 
of the inverse average wavenumber. 
(This choice for the set of  $\Delta \tau$ is made for mathematical convenience and generally does not strongly affect the statistical determination of whirl.) 
The first term in Eq.~(\ref{edisig:eqn}) represents the signal obtained with a conventional spectrograph while the second, cosine term
introduces a wavelength-dependent phase that provides additional leverage in measuring redshift.

The expected signal for one of the [OIII] lines of the SDSS target specified by Plate~\#2510, Fiber~\#560 is shown
in Figure~\ref{edicounts:fig}.   In this case the nominal delay is $\tau = 0.44$ mm with measurements
made in steps of 0.13 $\mu$m.

%%%%%%%%%%%%%%%%%%%%%%%%%%%% 
\begin{figure}[t]
   \centering
   \plotone{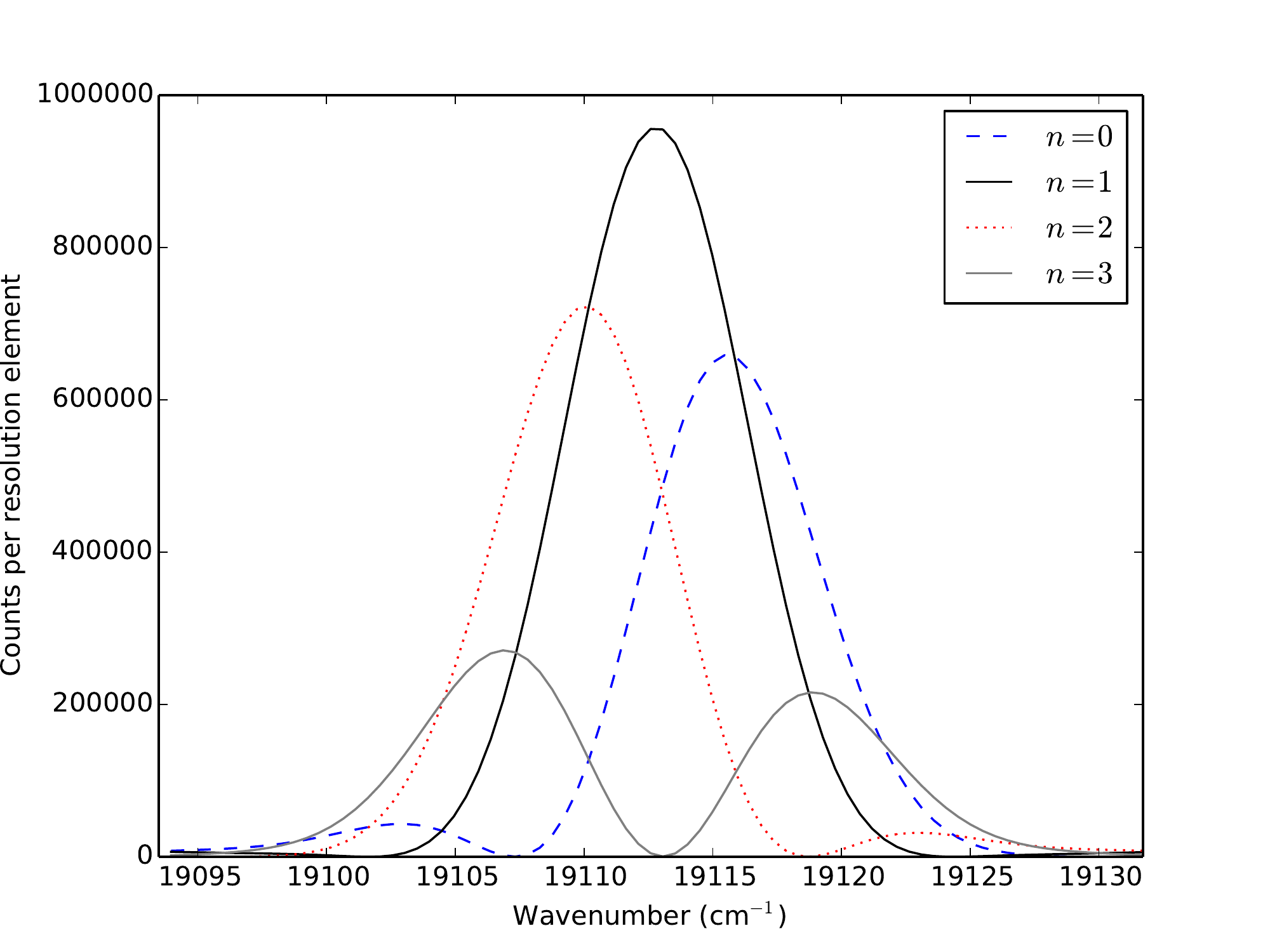} 
   \caption{The expected EDI signal from the four phase measurements for one of the [OIII] lines of
   the SDSS target specified by Plate~\#2510, Fiber~\#560. The four phases parametrized by $n$ are defined in Eq.~\ref{edisig:eqn}.
\label{edicounts:fig}}
\end{figure}

The redshift precisions on the target galaxies are given in Table~\ref{dz:tab}, where each of the four phases receives
a 2-hour exposure.  The EDI performance includes that of the Conventional spectrograph plus
the additional contribution from the whirl; in total the EDI (at $R=20,000$) outperforms the Conventional spectrograph (at $R=50,000$) 
in redshift precision by a factor of $\sim 3$. 

Separating the EDI information into Conventional and whirl contributions,
the signal and constraint from the Conventional spectrograph contribution shares the same systematic uncertainties as in Sec.~\ref{sec:dispspec}. 
The whirl component is best calibrated using arc lamps in the science exposure.  
The arc lines determine the configuration
of the EDI itself, $\phi_y$ and the $\Delta \tau$'s, and do not have to be close
to the science lines.  This avoids the temporal and wavelength interpolation needed in conventional spectrographs.
Wavelength uncertainty would then come from mismeasurement of phase and frequency due to pixels local to the
science lines, so absolute flux and PSF uncertainties become irrelevant. 

PSF calibration is facilitated with the presence of arc lines from which the
PSF  of the science lines can be interpolated.

%%%%%%%%%%%%%%%%%%%%%%%%%%%%%%%%%%% 
\subsection{Spatial Heterodyne Spectroscopy}
\label{SHS:sec} 

\subsubsection{Measuring Redshift}
Redshift can also be measured through the sum and the difference of observed wavenumbers of two lines whose restframe wavenumbers are known:
\begin{equation}
1+z=\frac{\sigma_{10}+\sigma_{20}}{\sigma_{1}+\sigma_{2}}=\frac{\sigma_{10}-\sigma_{20}}{\sigma_{1}-\sigma_{2}},
\label{redshift:eqn}
\end{equation}
where $\sigma_i$ is the observed and $\sigma_{i0}$ the restframe wavenumber of line $i$.
At face value, the measurement of $\sigma_1\pm\sigma_2$ should be expected to suffer larger uncertainty than that of the wavenumber
of a single line.  
To counter this expectation, we consider a scenario where the experimental setup has an output that is naturally sensitive
to $\sigma_1\pm\sigma_2$.  The differential measurement can be immune to certain systematic uncertainties encountered when using dispersion spectrographs.
Redshifts are measured from the relative shift of two lines of a doublet; 
the single line profile common to them enables cancellation of the PSF blurring of the
line profiles.

%%%%%%%%%%%%%%%%%%%%%%%%%%%%%%%%%% 
\subsubsection{Signal}

We consider Spatial Heterodyne Spectroscopy (SHS) \citep{1990SPIE.1235..622H}
as a means to get a direct measurement of  $\sigma_1\pm\sigma_2$.
Conceptually, the SHS can be understood as an interferometer with the  mirrors replaced by diffraction gratings.
(Practically we assume an interferometer configuration that  avoids the 50\% light loss inherent to  Michelson interferometers.)
The gratings have line-spacing $d$ and both are tilted
by $\theta$ with respect to the optical axis.  The deflected light from each grating exits the interferometer as a plane wave
propagating with angle $\gamma$ away from the optical axis,  given by the grating equation
\begin{equation}
\sigma\left(\sin{\theta}+\sin{\left(\theta-\gamma\right)}\right)=m/d,
\end{equation}
where $\sigma$ is the wavenumber and $m$ the diffraction order.
The angle of the emergent wave can be re-expressed as
\begin{align}
\sin{\gamma} & =-\cos{\theta} \left(\frac{m}{d\sigma} - \sin{\theta} \right) \nonumber \\
& \qquad + \sin{\theta}\sqrt{1-\left(\frac{m}{d\sigma} -\sin{\theta} \right)^2},
\end{align}
which is useful in calculations.

The two wavefronts from the two gratings, incident on the detector at angles $-\gamma$ and $\gamma$, interfere to make a fringe
pattern with  frequency
\begin{equation}
f_x=2\sigma\sin{\gamma}.
\end{equation}
For an input spectrum $B(\sigma)$, the intensity seen at position $x$ on the detector is
\begin{align}
I(x)& =\frac{1}{\Delta x}\left[\int_{0}^{\infty} B(\sigma)\left(1+\cos{\left(2 \pi x (2\sigma \sin{\gamma})\right)}\right)d\sigma \right. \nonumber \\
& \qquad \left.\vphantom{\int_{0}^{\infty}} \otimes \mbox{PSF}(x)\right]\Sha\left(\frac{x}{p}\right),
\end{align}
where  $p$ is the spacing of the wavenumber sampling, and for simplicity we assume a square collimated beam that covers
a range  $\Delta x \gg \left(2\sigma \sin{\gamma}\right)^{-1}$.
Since each spectral component is modulated by a distinct spatial
frequency at the output, the Fourier transform of $I(x)$ will recover the
input spectrum.
The Littrow wavenumber $\sigma_0$ is defined such that $2\sigma_0\sin{\theta}=m/d$, $\gamma=0$ and no fringe patterns are produced.

%%%%%%%%%%%%%%%%%%%%%%%%%%%%% 
\subsubsection{Toy Example}
\label{toy:sec} 

For illustration, consider a toy example where the input signal consists of two $\delta$-functions with the same intensity
\begin{equation}
B(\sigma)=\delta(\sigma_1)+\delta(\sigma_2).
\end{equation}
The output signal density in this case is
\begin{align}
I(x)&=\frac{1}{\Delta x}\left[2+ \cos{\left(2 \pi (2\sigma_1  \sin{\gamma_1})x \right)} \right. \nonumber \\
& \qquad \left.+ \cos{\left(2\pi (2  \sigma_2  \sin{\gamma_2})x \right)}\right]\\
&=\frac{2}{\Delta x}\left[1+ \cos{\left(2 \pi  (\sigma_1  \sin{\gamma_1}+ \sigma_2  \sin{\gamma_2})x \right)}  \right. \nonumber \\
& \qquad \left. \times\cos{\left(2\pi  (\sigma_1  \sin{\gamma_1}- \sigma_2  \sin{\gamma_2})x\right)}\right].
\end{align}
The two mixing of the sinusoidal outputs of the two lines results in two beat frequencies $(\sigma_1  \sin{\gamma_1}\pm\sigma_2  \sin{\gamma_2})$.

When the SHS is configured such that the Littrow wavenumber is
very close to $\sigma_1$ and $\sigma_2$ (and hence $\gamma_1$ and $\gamma_2$
are small),
\begin{equation}
\sigma_{1,2}\sin{\gamma_{1,2}} \approx  2(\sigma_{1,2}-\sigma_0) \tan{\theta}
\end{equation}
and the beat frequencies can be expressed as
$2(\sigma_1+\sigma_2-2\sigma_0)\tan{\theta}$ and $2(\sigma_1-\sigma_2)\tan{\theta}$.
Each beat frequency provides information on either 
$\sigma_1\pm \sigma_2$, the two quantities that directly lead to redshift.

The ratio $n$ between the beat frequencies can be chosen by adjusting the SHS configuration to give
\begin{equation}
\sigma_0=\frac{\sigma_1+\sigma_2}{2}-\frac{\sigma_1-\sigma_2}{2n}.
\label{littrowchoices:eqn}
\end{equation}
For $n=\infty$  the Littrow wavenumber is set to the average of the two lines: $\sigma_0=(\sigma_1+\sigma_2)/2$ and the first beat frequency is zero,
which in practice is difficult to quantify through measurement.  When $n=1$ the Littrow wavenumber is $\sigma_2$, meaning that line produces a flat signal and the other is the source
of the oscillations; the values of the two beat frequencies are equal.

%%%%%%%%%%%%%%%%%%%%%%%%%%%%%%%%%%%% 
\subsubsection{Performance on Target Galaxies}

We calculate the expected signal using the first order $m=1$ for a SHS with grating line density $1/d=1200$\,mm$^{-1}$
and using the more realistic double line model of Eq.~\ref{input:eqn}. 
Different configurations of the SHS are achieved by adjusting the grating tilt angle $\theta$, although swapping
gratings can achieve a similar effect.
The SHS configuration can then be expressed in terms of the Littrow wavenumber.

For the [OIII] doublet of the
SDSS target specified by Plate~\#2510, Fiber~\#560 
the condition for having only one beat frequency occurs when $\sigma_0=\sigma_2=25675$\,cm$^{-1}$, for 
$\theta=0.235870$  radians.
Different choices of Littrow wavenumber
produce different output signals:
Figure~\ref{shscounts:fig} shows the expected counts adjusting $\sigma_0$ to produce beat frequency ratios
$n=1/3.5,1/1.05,1,1.05,3.5$ according to Eq.~\ref{littrowchoices:eqn}.
The maximum $x$ is chosen to be $(12s_1\tan{\theta})^{-1}$ in order to cover the decay caused by the line-width.
The effect on the signal
of the choice of the beat frequency of $2(\sigma_1+\sigma_2-2\sigma_0)\tan{\theta}$ relative to the $2(\sigma_1-\sigma_2)\tan{\theta}$ frequency
is clearly apparent.
The precision is insensitive to the choice of $\sigma_0$, except
for small deviations near $n=1$ and at integer ratios of the beat frequencies; we choose
the case of $n=1.5$ or $\theta = 0.235900$  radians (a change in angle of $6.17 \arcsec$ from the $n=1$ case)
as a representative example where the pixels resolve the frequencies of interest.

%%%%%%%%%%%%%%%%%%%%%% 
\begin{figure}[t]
   \centering
   \plotone{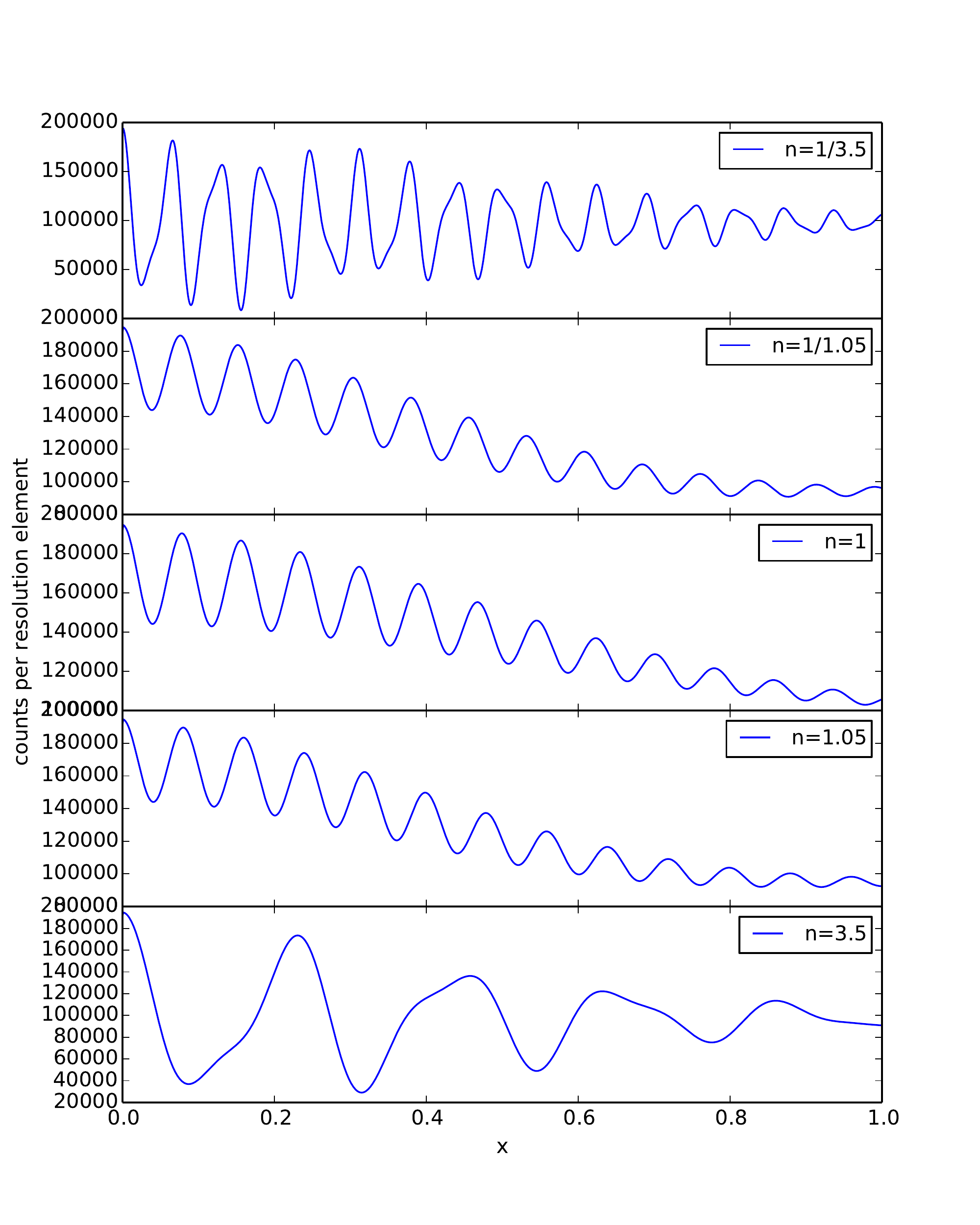} 
   \caption{The output signal for positive $x$ (it is symmetric about 
$x=0$) of five SHS configurations giving different
   Littrow numbers according to 
   $n=1/3.5, 1/1.05, 1, 1.05, 3.5$  in Eq.~\ref{littrowchoices:eqn}. 
   \label{shscounts:fig}}
\end{figure}

The redshift precisions on the target galaxies are given in Table~\ref{dz:tab}.  The statistical uncertainties are slightly worse than for
the Conventional spectrograph and well below EDI.  However, the results have reduced sensitivity to PSF uncertainties.

Wavelength calibration is tied to the determination of the SHS system parameters, the grating density $1/d$ and angle $\theta$.
Both can be calibrated externally through direct measurements, or arc lamp exposures; including lamp light
in science exposures is not preferred as all sources of light contribute the noise background.
Alternatively
the long exposures can be divided into sub-exposures each with a different SHS configuration, either by changing the
gratings or by rotating the grating angles by a series of $\delta \theta$.  The data themselves can be then used to fit for
the hardware parameters.

The line signal is distributed over thousands of pixels, reducing the pixel flux requirement to better than 1\%, much less stringent
than for the Conventional and EDI spectrographs.

Converting the spectrum to Fourier space transforms the sharp, imperfectly-known line profiles into the low-frequency envelope,
distinct from the high-frequency wiggles characteristic to each configuration that inform the redshift measurement.
The PSF blurring applies not in wavenumber space where it affects
line shapes but rather in physical space after the conversion of the spectral lines into cosine modulations. 

%%%%%%%%%%%%%%%%%%%%%%%%%%%%%% 
\subsection{ED-SHS} 

The two doublet lines combine to produce a mixed signal in an SHS.  Sending SHS light, such as those signals shown
in  Figure~\ref{shscounts:fig}, into an ensuing dispersion grating
allows the disentangling of the two signals for an unambiguous measurement of the frequency of each one.  Here, 
this scheme is 
referred to as an External Dispersion SHS (ED-SHS).
The observed signal is described by
\begin{align}
I(\sigma,x) &= \frac{1}{\Delta x}\left[\int^{\sigma \exp{\left(1/(2R)\right)}}_{\sigma} d\sigma'\,B(\sigma') \right.\nonumber\\ 
& \quad\qquad \times \left.\vphantom{\int_{0}^{\infty}} \left(1+\cos{\left(2 \pi x (2\sigma' \sin{\gamma})\right)}\right) \right] \nonumber\\ 
& \quad \otimes \mbox{PSF}(\sigma,x)\, \Sha\left(\frac{\sigma}{p_\sigma}\right) \Sha\left(\frac{x}{p_x}\right).
\label{edishssig:eqn}
\end{align}
The similarities between ED-SHS and EDI can be seen in Fig.~2 of \citet{2003PASP..115..255E}; both use
interferometry to create wavelength-dependent modulations in the signal, with the distinction
being that the EDI creates a uniform spatial frequency for all wavelengths and thus has an extremely wide bandwidth, whereas the ED-SHS creates a diamond-like fringe pattern whose spatial frequency varies rapidly around a specific 
wavelength, limiting the practical bandwidth significantly.

The redshift precisions on the target galaxies are given in Table~\ref{dz:tab}.  The statistical uncertainties are improved compared to those of the
SHS, and are comparable to the performance of the EDI.

Now that the signals from different wavelengths are no longer mixed, calibration arcs can be observed with the science image without increasing
the photon noise of the doublet lines.  Both wavelength and PSF calibration are simplified as with the EDI.
In addition, the flux signals per pixel are further reduced relative to the SHS, relaxing flux calibration requirements.

%%%%%%%%%%%%%%%%%%%%%%%%%%%%%%%%%%%%%%%%% 
\section{Conclusions} \label{sec:concl} 

Direct kinematic measurements of cosmic acceleration can provide valuable, 
model independent confirmation of dark energy. The accuracy required for 
detecting redshift drift is challenging, and we have reviewed some of the 
systematics, presented a series of theoretical speculations for observational 
probes, and explored various promising improvements to experimental 
approaches to measuring directly spectral line shifts with high accuracy.  

One result is that the greatest leverage on dark energy properties 
comes from low-redshift observations, due to lesser covariances there with 
other cosmological parameters. The dark energy figure of merit can exceed 
100 for a set of five measurements over $z=[0.1-0.5]$, each at 1\% precision 
(together with current precision on the Hubble constant), i.e.\ determining 
$w$ to 0.02, $w_0$ to 0.06, $w_a$ to 0.43. Moreover, redshift drift 
has extraordinary complementarity with other probes, having the opposite 
degeneracy in the $w_0$-$w_a$ plane from all standard probes. For example, 
combination of the low-redshift drift with Planck CMB data increases 
the figure of merit by a factor 12, to over 1000. Even a more practical 
5\% precision on redshift drift gives a FOM of 290 when combined with 
current Hubble constant and CMB data. Lesser complementarity can be achieved 
by combining redshift drift at low and high redshifts. 

Given the potential leverage of redshift drift, and the ability for low 
redshift measurements to be powerful, we reconsider the ``stare and wait'' 
method of observing spectral lines from a high redshift source (or a 
series of sources in the Lyman-$\alpha$ forest) over a time span of decades. 
We speculate about three different concepts. The first measures Hubble drift 
rather than redshift drift per se, gives a complementary measure of 
acceleration, and can be accomplished with normal galaxy redshift survey 
data using differential radial baryon acoustic oscillations (drBAO). 
The second would use cosmic ``pulsers'', not yet discovered, but possibly 
realizable through binary gravitational wave sources. Finally, long time-delay
strong gravitational lenses, possibly to be detected by future deep 
time domain surveys, are also considered. We emphasize that these ideas 
are speculative and need considerable development, but do have considerable 
synergy with surveys already planned such as DESI, Euclid, and LSST.  They 
can also replace patience -- of several decade time scales, see the 
categorization of \citet{stebbins} -- with ideas on how to get the universe 
and astrophysics to work for us. 

More practically, we turn to experimental approaches for improving high 
precision redshift measurements. Given the leverage of low redshift 
we focus on bright emission line galaxies as targets. We explore three 
alternatives to conventional spectrographs, involving differential and 
interferometric methods, and calculate their signal-to-noise 
and potential redshift precisions. We find that single source redshift 
measurement precision of $10^{-8}$ is potentially achievable, 
with further improvement possible from multiple sources, multinight 
exposures, and larger telescopes (our numbers are canonically for 8 hours 
on a 10-m telescope). ``Gold'' targets, with brighter and narrower doublet
emission than the galaxies considered in this article, may yet be discovered
in upcoming spectroscopic surveys and provide a speedier path toward
detecting redshift drift. 
Spatially resolving the galaxy, say with an integral field unit, may isolate
subregions with finer emission than that of the combined whole
\citep{2014arXiv1408.5765G}.
Instruments 
such as Externally Dispersed Interferometers and External Dispersion Spatial 
Heterodyne Spectroscopes can avoid some of the systematics in conventional 
spectrographs. 
Testing the stability and/or calibration precision possible on these 
instruments is of interest to see whether they can satisfy the requirements 
imposed by the science. 

If astrophysical systematics are high, interesting science may still come 
from peculiar acceleration maps, e.g.\ of the local Universe. 
Other targets, such as narrow radio lines or sequences of lines
from masers and molecular emission, should be 
explored, as should other uses of high-precision wavelength measurements, 
such as for atomic line catalogs \citep{2008PhLB..660...81A,
2008MNRAS.391.1308Q,
2012PhR...521...95Q}. 

Cosmic acceleration is a fundamental mystery of physics, and the possibility 
of adding a new probe of it -- one that is kinematic and does not depend on 
assumptions of how to separate matter and dark energy, or of their evolution -- 
is an exciting prospect. The hardware and systematics will be highly 
challenging, but success means that we will open up a new time domain: the 
ability to watch the universe as it evolves in real time.

%%%%%%%%%%%%%%%%%%%%%%%%%%%%%% 
\acknowledgments 

We acknowledge helpful discussions with Stephen Bailey, Jeff Newman, and 
Nao Suzuki. EVL thanks the 
Korea Astronomy and Space Science Institute for hospitality. 
This work has been supported by DOE grant DE-SC-0007867,
by LLNL under Contract DE-AC52-07NA27344, and the Director, 
Office of Science, Office of High Energy Physics, 
of the U.S.\ Department of Energy under Contract No.\ DE-AC02-05CH11231. 

Funding for SDSS-III has been provided by the Alfred P. Sloan Foundation, the Participating Institutions, the National Science Foundation, and the U.S. Department of Energy Office of Science. The SDSS-III web site is http://www.sdss3.org/.

SDSS-III is managed by the Astrophysical Research Consortium for the Participating Institutions of the SDSS-III Collaboration including the University of Arizona, the Brazilian Participation Group, Brookhaven National Laboratory, Carnegie Mellon University, University of Florida, the French Participation Group, the German Participation Group, Harvard University, the Instituto de Astrofisica de Canarias, the Michigan State/Notre Dame/JINA Participation Group, Johns Hopkins University, Lawrence Berkeley National Laboratory, Max Planck Institute for Astrophysics, Max Planck Institute for Extraterrestrial Physics, New Mexico State University, New York University, Ohio State University, Pennsylvania State University, University of Portsmouth, Princeton University, the Spanish Participation Group, University of Tokyo, University of Utah, Vanderbilt University, University of Virginia, University of Washington, and Yale University.

\bibliographystyle{hapj}
\bibliography{alex.bib}

\end{document}